\documentclass[1p,preprint]{elsarticle}

\journal{Physica A}

\begin{document}

\begin{frontmatter}

\title{Greenberg-Hastings dynamics on a small-world network: the effect of disorder on the collective extinct-active transition}

\author{Leonardo I. Reyes}
\address{School of Physics, Yachay Tech, Yachay City of Knowledge 100119-Urcuqu\'{\i}, Ecuador\\and\\
Departamento de F\'{\i}sica, Universidad Sim\'on Bol\'{\i}var, Apartado Postal 89000, Caracas 1080-A, Venezuela}
\ead{lreyes@yachaytech.edu.ec}

\begin{abstract}
We present a numerical study of a reaction-diffusion model on a small-world network. 
We characterize the model's average activity $F_T$ after $T$ time steps and the transition
from a collective (global) extinct state to an active state in parameter space.
We provide an explicit relation between the parameters of our model at the frontier between these states.
A collective active state can be associated to a global epidemic spread, or to a persistent neuronal activity.
We found that $F_T$ does not depends on disorder in the network if the transmission rate $r$
or the average coordination number $K$ are large enough.
The collective extinct-active transition can be induced by changing two parameters
associated to the network: $K$ and the disorder parameter $p$ (which controls the variance of $K$).
We can also induce the transition by changing $r$, which controls the threshold size in the dynamics.
In order to operate at the transition the parameters of the model must satisfy the relation $rK=a_p$,
where $a_p$ as a function of $p/(1-p)$ is a stretched exponential function. 
Our results are relevant for systems that operate {\it at} the transition in order to 
increase its dynamic range and/or to operate under optimal information-processing conditions.
We discuss how glassy behaviour appears within our model. 
\end{abstract}


\end{frontmatter}

Many problems in Science can be cast in terms of dynamics on networks: social phenomena \cite{grano1,mytho,weightedNet}, 
 epidemic spread \cite{dickman}, food webs \cite{lazlo} and ecosystem's diversity \cite{preEco},
brain activity \cite{prlBrain,prlAvala,physicaA,SWbrain,schizo,mehta}, granular materials \cite{granular1,granular2,radjai} and, in general, complex systems \cite{ottino}. Among the most studied models, the small-world network model of Watts and Strogatz (WS) \cite{natureWS,SW2000,SWbrain} can be tuned to interpolate between a regular and a random network, a very atractive property that allows us to explore the consequences of network disorder on dynamics. In this work we consider a stochastic reaction-difusion cellular automata model 
on a small-world network and study its average activity after $T$ time steps and its collective extinct-active transition. 
For the first time, we provide a explicit relation between the parameters of the model for the system to operate {\it at} the transition,
and disorder enters in this relation as a stretched exponential function.

In the WS model we start from an ordered ring in which each node is conected to its $K$ nearest neighbours, and then we randomly rewire each conexion with probability $p$. Since we only rewire and we don't create new conexions, the average coordination number is still $K$, but
the distribution of $K$ gets broader as we increase $p$ \cite{SW2000}.
Two quantities typically used to characterize a network are the mean minimal distance between two nodes $L$
and the clustering coeficient $C$. For each node, we can measure how many of it's neighbours are conected between them. $C$ is defined as the actual number of conexions between a node's neighbours over the maximun number of posible conexions between neighbours, averaged over all nodes. Hence, in a social network $C$ could measure up to what extent someone friends are friends between them. In the WS network there is a range of values of $p$ for which $C$ is still high and $L$ is small, a signature of a small-world scenario. 
The case $p=0$ corresponds to an ordered network, with high $C$ and $L$. 
The case $p=1$ corresponds to a random network, with small $C$ and $L$ \cite{SW2000,natureWS}. 

A very generic reaction-difusion model is the Greenberg-Hastings model (GH). 
This model has been used to model the Belousov chemical reaction, biological systems, epidemic spread, collective behaviour of heart cells and neuronal activity  \cite{GH1,GH,MathBio,BerryFates,polonia,global2005,pre,prlBrain}. 
Our stochastic version of this three state model is as follows:

\begin{itemize}
\item If a cell is in the {\it excited} state at time $t$ then it is in the {\it passive} state at time $t+1$.

\item If a cell is in the passive state at time $t$ then it is in the {\it susceptible} state at time $t+1$.

\item If a given cell is in the susceptible state at time $t$, and at least one of its neighbours is in the excited state at time $t$, then the given cell is in the excited state at time $t+1$ with probability $r$, otherwise the given cell remains in the susceptible state.
\end{itemize}

The GH is a model for excitable media (\cite{ott}, see fig. 1 in \cite{excitableDyn}) whose threshold is controlled by the parameter $r$. We call the parameter $r$ the transmission (or infection) probability. If {\it excited} means no-healthy, then I become infected if at least one of my neighbours is infected, but with probability $r$. In this context $r$ would be a measure of the average state of the inmune system of the population, with a smaller $r$ implying
more resistance to become ill. If we are in a social context, $r$ could be associated to the confidence of the social agents on an especific action, with greater $r$ implying more confidence. For neuronal activity or chemical reactions, $r$ can be associated to a potential barrier, with a smaller $r$ implying a greater potencial barrier (a greater threshold).
In a raw model for dense granular flows, the states of the GH model could be associated to 
fluid or solid type of contacts between grains \cite{aranson}, with the excited state associated to fluid contacts,
the passive state associated to contacts with static friction and with a large time of contact \cite{aranson}
or with a small mobilization of friction \cite{radjai}, and
the susceptible state associated to contacts with 
static friction and with a short time of contact or with a large mobilization of friction.
Versions of the GH model have been implemented on regular networks \cite{assis,BerryFates,global2005}, scale free networks \cite{pre},
on a small-world network to study spiral waves \cite{sinica} and on the human connectome \cite{prlBrain}.

\begin{figure}
\includegraphics[width=0.99\textwidth]{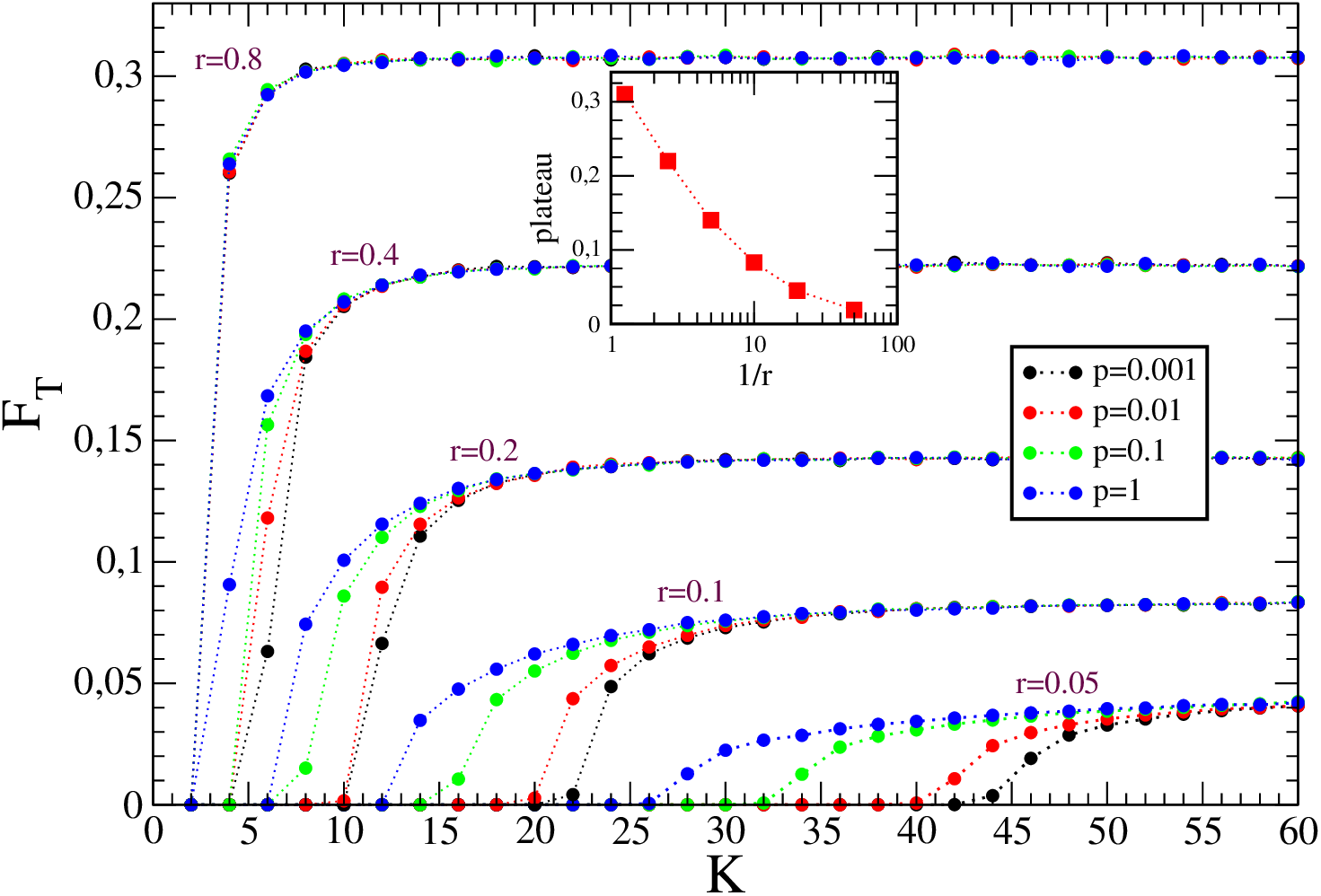}	
\caption[global1]
{Average activity $F_T$ as a function of the average coordination number $K$, for several values of $p$ and $r$. 
In the inset is shown the plateau of $F_T$, obtained for large $K$, as a function of $1/r$.
For $r=0.05$ and $p=0.1$, for example, we have a collective active state for $K>32$. If $r\rightarrow 1$ or if $K$ is large enough
$F_T$ becomes independent of the disorder parameter $p$. 
The number of nodes is $N=1000$ and we show the average result for $100$ realizations for each set of parameters. 
For this figure, we looked for activity in the system after $T=1000$ time steps.
\label{global1}}
\end{figure}

It is of particular interest under what conditions the system, for a given initial condition, evolves towards a global active state, in which a finite fraction of the nodes remains active \cite{BerryFates}. A collective active state can be associated to a global epidemic spread, or to a persistent neuronal activity. For random initial conditions, we have found transitions to a collective active state as we vary any of the three parameteres in the GHWS model: the transition can be induced
by increasing the average coordination number $K$, 
by increasing the disorder in the network $p$ or by increasing the transmission probability $r$. 
Based on our numerical results, we provide for the first time a 
explicit relation between the parameters of the model for the system to operate {\it at} the transition,
and disorder enters this relation as a stretched exponential function. 
Stretched exponentials are usually found in glasses and disordered systems \cite{stretch1,stretch2,stretch3}.
As far as we know, for this model (GHWS) \cite{sinica} we are presenting a novel way to consider the influence of disorder on 
the extinct-active frontier. In reference \cite{prlBrain} the network was the human connectome and only a threshold was varied 
in order to tune the transition.

Our system consists of $N=1000$ nodes.
Initially, each node is randomly asigned, with equal probability, to one of the three states of the model.
We call $F$ the average number of nodes in the excited state.
The behaviour of the GHWS model is sumarized in figure \ref{global1}, on which we show the average activity
after $T=1000$ time steps $F_T(K,p,r)$.  
We have a collective active state if after $T$ time steps (see below) $F_T>0$.
We have a collective extinct state if after $T$ time steps $F_T=0$.
For $K>K_c$ we have a collective active state,
and from this figure we see that $K_c=K_c(r,p)$. 
It can be observed that for $r\rightarrow 1$ $F_T$ becomes independent of the disorder parameter $p$.
Also, for any given transmission probability $r$, $F_T$ becomes independent of $p$ if $K$ is
large enough: it tends to a plateau whose dependence on $1/r$ is shown in the inset. 
 
\begin{figure}
\includegraphics[width=0.99\textwidth]{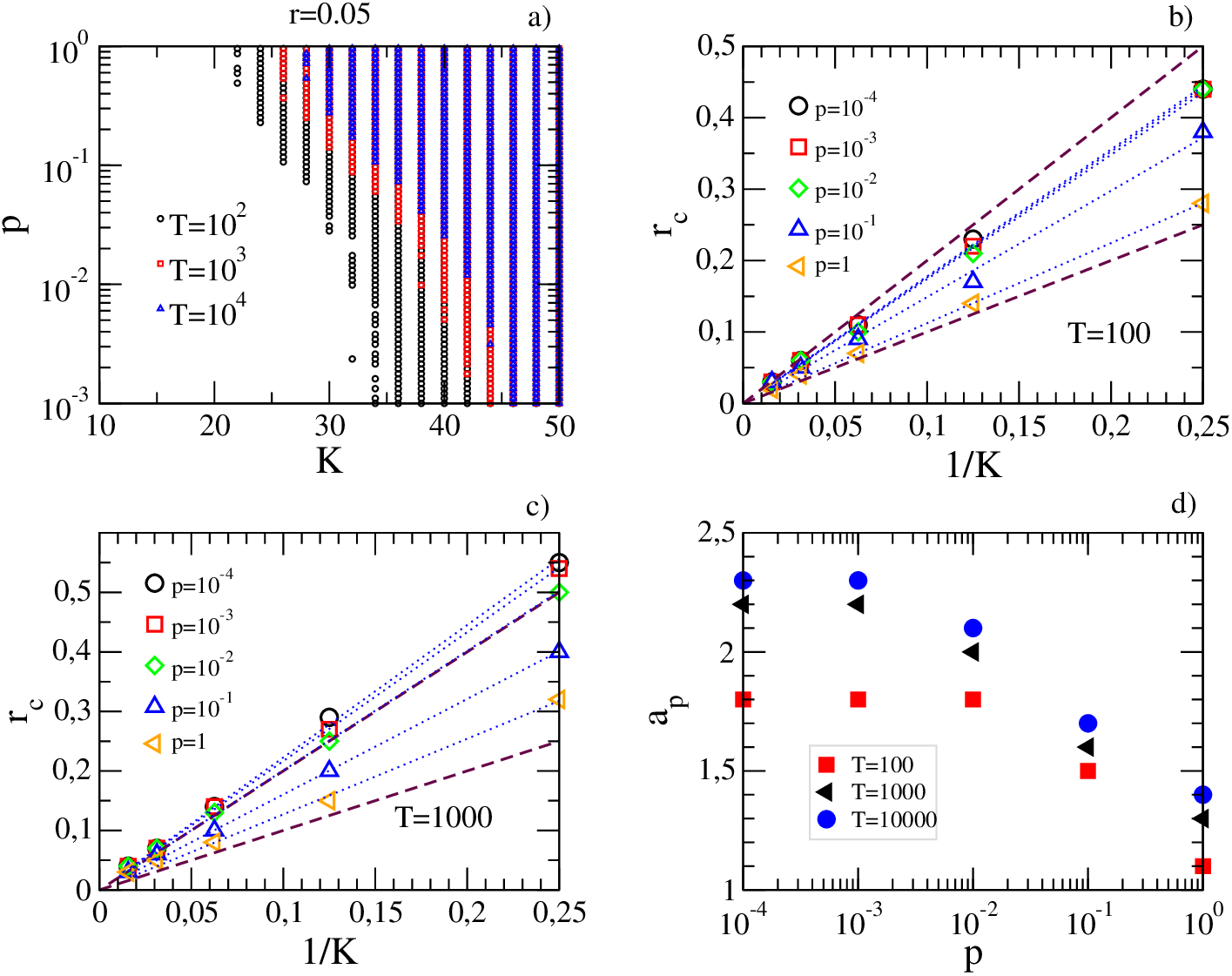}	
\caption[rcfig]
{a) Zones of collective active states for $r=0.05$ after $T$ time steps.
b) Critical transmission probability $r_c$ as a function of $1/K$, 
estimated after $T=100$ time steps for different $p$.
c) Same as in b) but with $T=1000$. As in b), the dashed lines corresponds to the relations $r_c=1/K$ and $r_c=2/K$ .
Our system is a network with $N=1000$ nodes and we averaged over $m=100$ realizations.
Dotted lines, here and in b), are best fits to the relation $r_c=a_p/K$. 
We don't considered the case $r=1$ (the deterministic GH model)
for which we have a collective active state for $K=2$, for any $p$. 
d) The slope $a_p$ as a function of the disorder parameter $p$, obtained from c) and b).
\label{rcfig}}
\end{figure}

For $r=0.05$, we show in figure \ref{rcfig}a zones of global active states in parameter space.
We have found, as previously reported \cite{goltsev}, that for $K<K_c$
the relaxation time of decay to zero activity increases largely as we approach a transition to a collective active state.
Because of this, we looked for activity after $T$ time steps, and by a {\it time step} we mean an update of the whole network.
Thus, we are characterizing the evolution in time of the frontier, and the limit $T\rightarrow\infty$ corresponds to the usual
meaning of the frontier. 
For a given $r$, we can induce a transition to a collective active state by increasing the disorder parameter $p$ or 
by increasing the average coordination number $K$.

In order to characterize the frontier between active and extinct global states in parameter space we show in
figures \ref{rcfig}b and  \ref{rcfig}c the critical transmission probability $r_c$ as a function of $1/K$, 
for disorder spanning four orders of magnitude in $p$. 
Our results can be sumarized in the following simple relation:
\begin{equation}\label{eq1}
r_c=\frac{a_p}{K}.
\end{equation}

This scaling of $r_c$ with $K$ was obtained by Berry and Fates \cite{BerryFates} in a mean field approximation of a similar GH model,
where the crucial ingredient was to obtain an approximation to the conditional probability that in the neighbourhood of
a given node there is at least one node in the excited state, given that the considered node is in the susceptible state.
The scaling $r_c\sim K^{-1}$ was verified by numerical simulations 
on a regular network by considering different types of neighbourhoods, obtaining $a_p\approx 2$ 
(the exact mean field result was $a_p=1$) \cite{BerryFates};
this result was robust against the inclusion of defects in the regular network.
In figure \ref{rcfig}d we show $a_p$ as a function of $p$, and
we conclude from this figure that disorder in the network reduces $r_c$. Since we have a collective active state for $r>r_c$, 
within the GHWS model disorder favours a collective active state.

\begin{figure}
\includegraphics[width=0.99\textwidth]{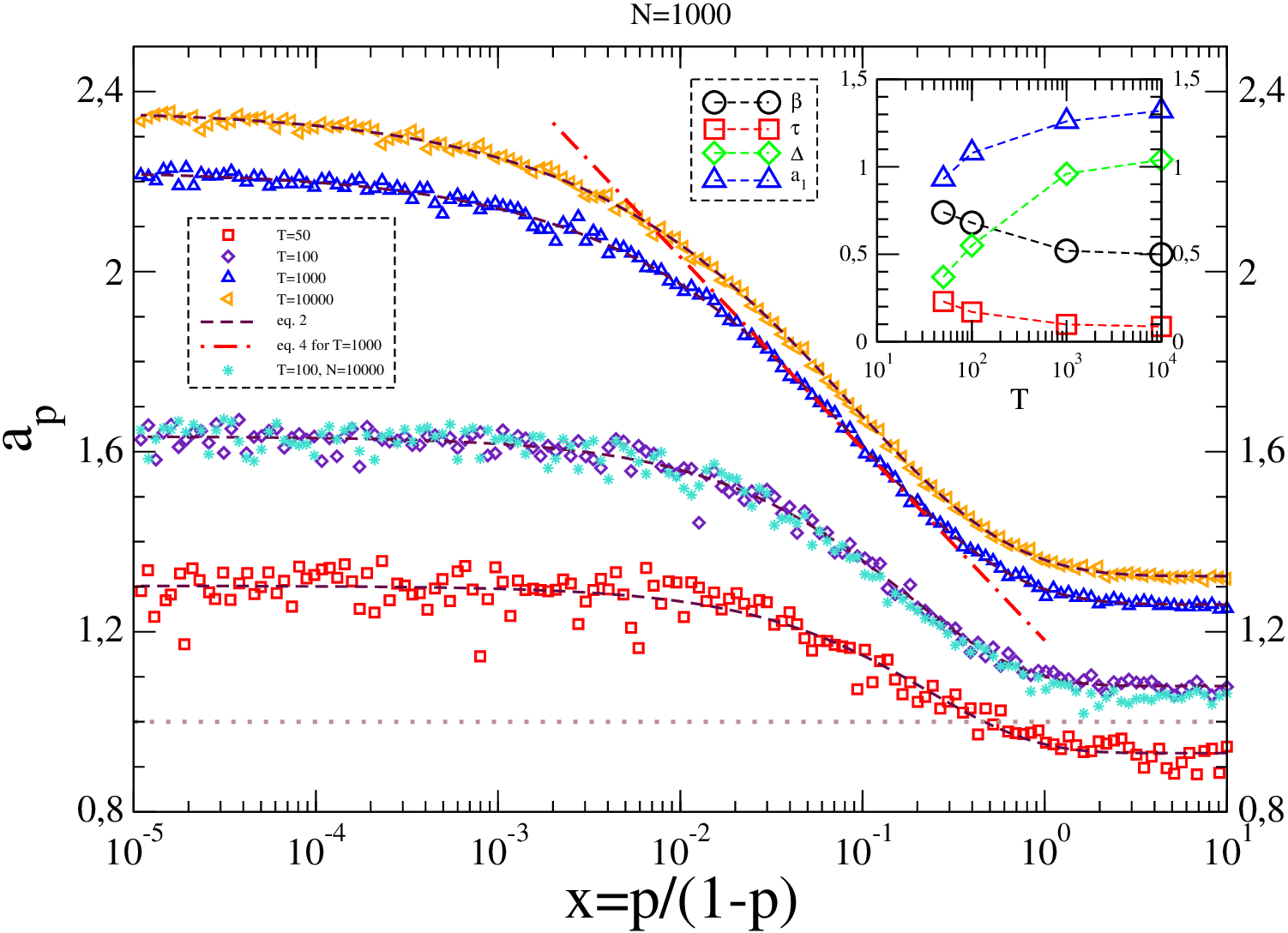}	
\caption[fig3]
{This figure is a detailed version of figure \ref{rcfig}d. 
In order to obtain each point in this graph we considered for each $p$ five values of $K$: $K=4,8,16,32,64$,
with $m$ realizations for each $K$. We then adjusted the best line to equation (\ref{eq1}).
For $T=50$ and $T=100$, we used $m=1000$; for $T=1000$ we used $m=100$; for $T=10000$ we used $m=30$.
Dashed lines are best fits to stretched exponentials of the form given by equation \ref{eqSt}.
In the inset we show the four parameters of eq. \ref{eqSt} as a function of $T$. 
The dot-dashed line is equation \ref{eq3} for $T=1000$. 
We show, in stars, the results for $N=10000$ and $T=100$, with no major differences with the case $N=1000$;
certainly $L$ depends on $N$ \cite{SW2000} 
but $L$ appears to affect only the time required to propagate the activity to the whole system \cite{natureWS}. 
The dotted line is the mean field result $a_p=1$ \cite{BerryFates}.
Recently (v6), I have been aware that if we average in a different way we do have a $N$ dependence: if we do only
one realization for each value of $K$ (see the beginning of this caption), obtaining $a_p$ and then average several
realizations we do have a $N$ dependende in the results. This fact will be considered elsewhere. For $N=1000$, the results
presented here do not change qualitatively.

\label{fig3}}
\end{figure}

In figure \ref{fig3} we see a detailed version of figure \ref{rcfig}d. Our numerical results for $a_p$ 
 are consistent with a stretched exponential of the form:
\begin{equation}\label{eqSt}
a_p=\Delta\exp[-(x/\tau)^\beta]+a_1
\end{equation}
with $x=p/(1-p)$. Stretched exponentials are usually found in relaxation of glasses and 
disordered systems \cite{stretch1,stretch2,stretch3,stretch4,stretch5}.
In the inset of figure \ref{fig3} we show the evolution of the four parameters of equation \ref{eqSt}. 

In terms of $z\equiv \ln(1/x)$, there is a maximum in the quantity $\chi(z)\equiv\partial a_p/\partial z$ at $z_c=\ln(1/\tau)$, with
the maximum given by $\chi_m\equiv\chi(z_c)$:
\begin{equation}
\chi_m=\frac{\Delta\beta}{e}
\end{equation}

\begin{figure}
\includegraphics[width=0.99\textwidth]{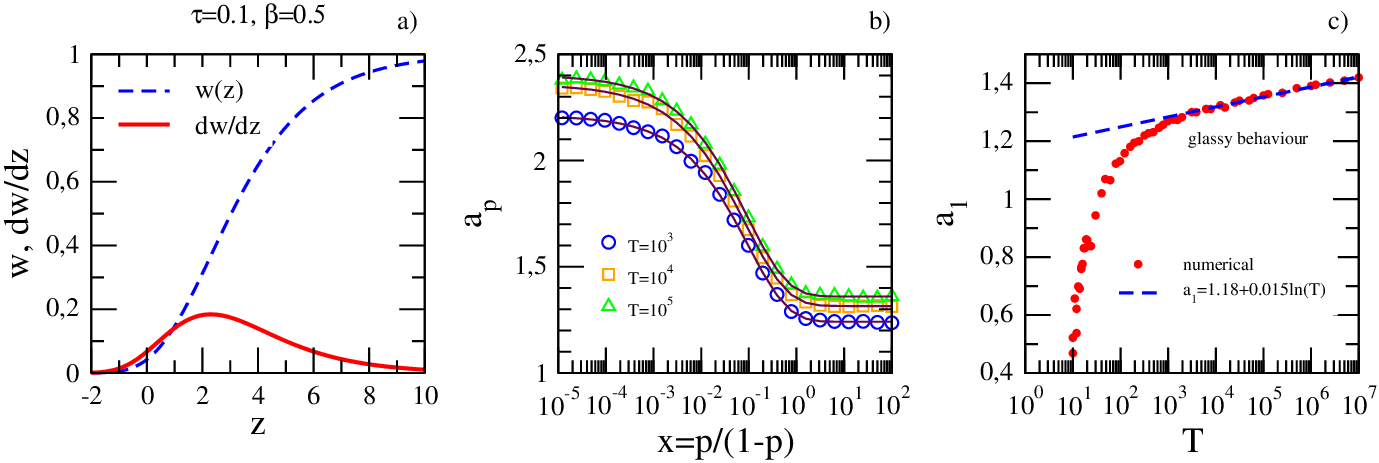}	
\caption[fig4]
{a) Behaviour of $w\equiv (a_p-a_1)/\Delta$, and it's derivative, with $\chi=\Delta\partial w/\partial z$.
b) $a_p$ for $T=10^3$, $T=10^4$ and $T=10^5$ (different run from previous figure).
Dashed lines are best fits to stretched exponentials of the form given by eq. (\ref{eqSt}).
At $T~10^4$ the frontier no longer changes it's shape,
all the parameters of the stretched exponential no longer changes with time, except for $a_1$.
For $rK<a_p$ we have an extinct state.
c) The frontier keeps moving since $a_1$ keeps increasing (as $a_1\sim 0.015\ln(T)$ for $T>2000$). 
We could say that for $T>2000$ the frontier enters a glassy-like regime. 
\label{fig4}}
\end{figure}

The quantity $\chi$ measures the sensitivity of the extinct-active frontier to changes
in disorder (through the variable $z$), going to zero for very ordered or very disordered networks.
In terms of $\mu\equiv e^{-z}/\tau$, the order parameter $w\equiv (a_p-a_1)/\Delta=\exp({-\mu^\beta})$ and 
$\chi(z)=\Delta\partial w/\partial z$, we have: $\chi(z)/\chi_m=e\mu w$ (see figure \ref{fig4}a).
If we operate at $z_c$ we can communicate more easily through collective states when considering, for example, a network of networks 
\cite{netofnets} in which each unit is a GHWS one: 
at the frontier and at $z_c$ small changes in disorder can allow the system to switch between an extinct and
an active collective state more easily.

If we expand around $z_c$, we obtain the approximate relation:
\begin{equation}\label{eq3}
a_p-a_{pc}\approx \chi_m (z-z_c)
\end{equation}
with $a_{pc}=\Delta/e+a_1$. In figure \ref{fig3} we show eq. \ref{eq3} for $T=1000$.

By monitoring the activity $F$ as a function of $T$ we have found that $a_p$ is well 
described by a stretched exponential function in the variable $x=p/(1-p)$, for all $T$. 
Since at the extinct-active frontier $rK=a_p$ (eq. \ref{eq1}), we can interpret $a_p$ as the average minimal
number of {\it effective} neighbours to become active. In eq. \ref{eqSt} we have defined $\Delta$ as the difference between $a_0$
($a_p$ when $p\rightarrow 0$) and $a_1$ ($a_p$ when $p\rightarrow 1$), see the inset of figure \ref{fig3}. 
What happens, for example, if we consider an ensamble of systems such that $rK\approx 1.63$, 
in which each member of the ensamble can have any value of $p$?. We see in figure \ref{fig3} that $a_0\approx 1.63$ for $T=100$.
Thus, up to $T=100$ we'll have ordered and disordered members of the ensamble that are still active,
but in the long run only the more disordered members of the ensamble ($x>0.2$, approx., see fig. \ref{fig3}) will stay active.
If we consider an ensemble of systems with $rK<a_1$, then in the long run all the members of the ensemble will go extinct.
On the other hand, if we consider an ensemble of systems with $rK>a_o$ 
then all the members of the ensemble will stay active in the long run.

In figure \ref{fig4}b we can see that the frontier no longer changes it's shape for $T>10^4$, only $a_1$ keeps increasing
with time, as can be seen in figure \ref{fig4}c. 
With the relation $a_1=1.18+0.015\ln(T)$ being satisfied for $T>2000$ (see fig. \ref{fig4}c), we have
that for an ensamble with $rK\approx 2.4$, at $T_0=10^5$ the more ordered members of the ensamble starts to go extinct
(see fig. \ref{fig4}b).
For this ensamble, if we denote $T_1$ as the time when the more disordered members of the ensamble will go extinct we have
that $T_1/T_0=e^{\Delta/0.015}\approx e^{67}\approx 10^{29}$. This is reminiscent of glassy behaviour.

Several generalizations can be introduced in the model implemented here. One of them would be to consider a weighted 
network, with weights that can be correlated or not to the local coordination number \cite{weightedNet}.  
In order to consider different passive time scales, a possible generalization of our GH model would
be to introduce $M$ time steps in the passive (refractory) state before becoming susceptible. Regarding this possibility,
 in reference \cite{BerryFates} was found that, for a regular network, 
the critical transmission probability depends weakly on $M$, and that 
{\it at} the active-extinct transition the decay in time of the average activity $F$ follows a power law, with 
an exponent that does not depends on $M$: its universality class remains directed percolation.

We have characterized the activity of a GHWS model, and we found that
the average activity does not depends on disorder in the network if the transmission rate $r$
or the average coordination number $K$ are large enough.
We have found that a collective extinct-active transition in a GHWS model can be induced by increasing $r$,
a parameter that controls the threshold size in the system's dynamics. 
Also, the transition can be induced by increasing the disorder parameter
$p$ and by increasing $K$. Our results are relevant to systems that need to operate {\it at} the 
extinct-active transition, in order to increse its dynamic range and/or to operate under optimal information-processing conditions 
\cite{natureOptimal}.
Remarkably, in reference \cite{prlBrain} it was found that in order to reproduce the patterns of neuronal activity observed in experiments,
the threshold of a stochastic GH model implemented on the human connectome has to be tuned for the system to operate at the 
the extinct-active transition \cite{beingCritical}. 
Within the GHWS model studied here, 
in order to operate at the transition the parameters of the model must satisfy
the relation $rK=a_p$, where $a_p$ as a function of $p/(1-p)$ is a stretched exponential function.

I would like to acknowledge Miguel Pineda for usefull discussions
during the first stages of this investigation.


\end{document}